\documentclass[11pt, a4paper]{article}
\pdfoutput=1
\usepackage{graphicx}
\usepackage{jcappub}

\usepackage{graphicx}
\usepackage{dcolumn}
\usepackage{bm}
\usepackage{subfigure}

\newcommand{\lyaf}{Ly$\alpha$ forest}
\newcommand{\bF}{\bar{F}}

\newcommand{\vx}{\mathbf{x}}

\begin{document}

\title{Characterizing the \lyaf\ flux probability distribution function using Legendre
  polynomials}

\author{Agnieszka M. Cieplak,}
\emailAdd{acieplak@bnl.gov}
\affiliation{Brookhaven National Laboratory, \\
             Bldg 510, Upton NY 11973, USA}

\author{An\v{z}e Slosar}
\emailAdd{anze@bnl.gov}

\date{\today}

\abstract{ The Lyman-$\alpha$ forest is a highly non-linear field with
  considerable information available in the data beyond the
  power spectrum. The flux probability distribution function (PDF) has
  been used as a successful probe of small-scale physics. In this
  paper we argue that measuring coefficients of the Legendre
  polynomial expansion of the PDF offers several advantages over
  measuring the binned values as is commonly done. In particular, the
  $n$-th Legendre coefficient can be expressed as a linear combination
  of the first $n$ moments, allowing these coefficients to be measured
  in the presence of noise and allowing a clear route for
  marginalisation over mean flux. Moreover, in the presence of noise,
  our numerical work shows that a finite number of coefficients are
  well measured with a very sharp transition into noise
  dominance. This compresses the available information into a small
  number of well-measured quantities. We find that the amount of
  recoverable information is a very non-linear function of spectral
  noise that strongly favors fewer quasars measured at better signal
  to noise.}


\maketitle

\section{Introduction}

The \lyaf, seen as absorption lines of neutral hydrogen along the
line-of-sight in quasar spectra, has become a powerful tracer of the
underlying matter distribution at high redshift. Its multiple
statistics \cite{2000ApJ...543....1M}, ranging from the evolution of the
mean flux with redshift, the transmitted flux probability distribution
function (PDF), to the two-point statistics of the power spectrum and
correlation function, have been successfully used to probe
cosmological parameters (as seen in \cite{2013PhRvD..88d3502V,
  2006PhRvL..97s1303S, 2003ApJ...594L..71A, 2015JCAP...02..045P, LESGOURGUES2015, 2006JCAP...10..014S, 2006JCAP...10..014S, 2014PhRvD..89b3519D, 2013A&A...552A..96B, 2013JCAP...04..026S,
  2015A&A...574A..59D} for example).

The flux PDF in particular has been studied since  \cite{1991ApJ...376...33J} as a way to understand the amplitude of matter fluctuations and
the thermal history of the intergalactic medium (IGM). There has been
much renewed interest in the flux PDF, since \cite{2008MNRAS.386.1131B} and \cite{2009MNRAS.399L..39V} have found a possible inverted temperature-density
relation at $2<z<3$ with high resolution, high signal-to-noise (S/N)
spectra. This would be in contrast to the theoretical predictions of a
positive relation of $T \sim \rho^\gamma$, with $\gamma \sim
1.6$.
Most recently, \cite{2015ApJ...799..196L} revisited this question with
stacked low resolution Sloan Digital Sky Survey (SDSS) spectra and found no such inversion. The
flux PDF is therefore an ongoing probe of interest.

However, in addition to being sensitive to the cosmological evolution
above, it is also dependent on the pixel noise level, the resolution
of the spectra, and the systematic uncertainties in fitting the
continuum level \cite{2015ApJ...799..196L}. Therefore, the proper
treatment of these errors is crucial. \cite{2000ApJ...543....1M} first
demonstrated the importance of using covariance matrices instead of
the pure diagonal error bars in flux PDF measurements. The problem
arises with the inversion of this flux PDF covariance matrix, as it is
exactly singular \cite{2006ApJ...638...27L} due to the normalization
integral constraint, making it unnecessarily difficult to calculate
noise and correctly fit predicted models \cite{2015ApJ...799..196L}.

Here we present an alternative way of encoding the same information,
which is arguably systematically cleaner and conceptually nicer. 

\section{Shifted Legendre polynomial expansion}

The flux field is a three-dimensional field defined through the 3D space
\begin{equation}
  F(\vx) = e^{-\tau(\vx)} = \bar{F}\left(1+\delta_F(\vx) \right),
\end{equation}
where $\tau(\vx)$ is the optical depth due to the \lyaf, $\bar{F}$ is
the mean flux and $\delta_F$ are variations around this mean flux. The
statistical properties of this field are completely described by the
full set of $n$-correlation functions
$\left<F(\vx) \cdots F(\vx') \right>$. Usually we measure the 2-point
function in real or Fourier space, but considerable information is present in
the field beyond these two. \cite{2003MNRAS.344..776M} demonstrates
that the 1D bispectrum contains at least the same amount of
information as the 1D power spectrum. Alternatively, and perhaps somewhat
easier routes are via PDFs. After optionally smoothing the field on a
certain scale (e.g., due to the instrumental resolution of a spectrograph), the flux
PDF $P(F)dF$ is defined as the probability of the flux being between $F$
and $F+dF$ for a randomly chosen point. This function is uniquely
determinable from all possible cumulants (which are in turn given by
moments) of the input field and of course also depends on
smoothing. 

The zeroth moment of the field is unity and the first moment gives the
mean flux:
\begin{eqnarray}
  \int_0^1 P(F) dF &=& 1,\\
  \int_0^1 FP(F) dF &=& \bar{F}.
\end{eqnarray}

Traditionally, the flux PDF has been measured in $i=1 \ldots N$ bins of flux, defined
by 
\begin{equation}
  B_i = \int_{(i-1)/N}^{i/N} P(F) dF
\end{equation}
This gives the constraint $\sum B_i=1$. While one can write a Bayesian
estimator for $B_i$, it is clear that at least the covariance matrix
of the measured $B_i$ will be non-positive definite, since the
linear combination of $\sum B_i$ modes has zero variance and hence the
corresponding eigenvalue is zero, making the covariance matrix
strictly non-positive definite. There are robust ways of
  dealing with this. For example, one can diagonalize the matrix and use
  the $N-1$ non-zero eigenvectors to determine
  $\chi^2$. Alternatively, one can simply drop one of the bins (since
  its value is completely determined by the sum constraint). However,
  while matrix inversion is a nuissance, there are two more significant
problems. First, in the presence of noise, a simple binning of flux
will not give an unbiased estimate of $B_i$. Second, unless one deals
with very high signal-to-noise ratio (SNR) data measured at very high resolution, we can only
measure values of $\delta_F$ rather than $F$ and hence some
marginalisation over $\bar{F}$ is required.

Here we propose an alternative, namely to expand $P(F)$ in terms of
shifted Legendre polynomials $L_\ell(x)=\tilde{L}_\ell(2x-1)$, where
$\tilde{L}_\ell$ are the standard, unshifted polynomials defined
between $-1$ and $+1$ and $L_\ell(F)$ are the shifted version defined
between $0$ and $1$. Given that this is a complete orthonormal basis, we have:

\begin{eqnarray}
P(F) &=& \sum_{\ell=0}^{\infty}a_\ell L_\ell(F),\\
a_\ell &=& (2\ell +1) \int_0^1 P(F) L_\ell(F) dF
\end{eqnarray}
The first few polynomials are given by 
\begin{eqnarray}
  L_0 (F) &=& 1 \\
  L_1 (F) &=& 2F-1 \\
  L_2 (F) &=& 6F^2-6F+1 \\
  L_3 (F) &=& 20F^3-30F^2+12F-1\\
  L_4 (F) &=& 70F^4 - 140F^3+90F^2-20F+1\\
&\cdots&
\end{eqnarray}
Using standard properties of Legendre polynomials, one 
finds that moments of the flux field are given by
\begin{equation}
M_n = \left< F^n \right> = \sum_{l=0}^n \frac{a_\ell n!^2}{(n-l)!(n+l+1)!}.
\label{eq:mom1}
\end{equation}
This result is crucial since we have made an explicit link
between the moments of the distribution and the PDF. More importantly,
we have also shown that the first $n$
coefficients $a_\ell$ are completely determined by the first $n$
moments $M_n$.

These can be converted to the more observationally relevant quantity: the moments of $\delta_F$,
given by $m_n=\left<\delta_F^n\right>$, with $m_0=1$, $m_1=0$. Here we have
\begin{equation}
M_n = \bar{F}^n \sum_{k=0}^n {{n}\choose{k}}  m_{n-k}.
\end{equation}

We can therefore express the coefficients of our polynomial expansion $a_\ell$
combinatorially from observed moments $M_n$ or $m_n$ as
\begin{eqnarray}
  a_0 &=& 1  \\
  a_1 &=& 6M_1 -3 \\
  a_2 &=& 30M_2-30M_1+5 = 30\bF^2 m_2+30\bF^2-30\bF+5 \\
  a_3 &=& 140M_3-210M_2+84M_1-7 \\&=&140\,{\bF}^{3}{m_{
{3}}}+420\,{\bF}^{3}{m_{{2}}}-
210\,{\bF}^{2}{m_{{2}}}+140\,{\bF}^{3}
-210\,{\bF}^{2} +84\,\bF-7  \nonumber \\
 a_4 &=& 630M_4-1260M_3+810M_2-180M_1+9 \\&=&630\,{\bF}^{4}{m_{{4}}}+2520\,{\bF}^{4}{m_{{3}}}-1260\,{\bF}^{3}{m_{{3}}} +3780 {\bF}^{4}{m_2}-3780\,{\bF}^{3}{m_{{2}}} \nonumber \\&&+810\,{\bF}^{2}{m_{{2}}}+630 {\bF}^4-1260\,{\bF}^{3}+810\,{\bF}^{2} -180\,\bF+9 \nonumber
  \\
&\cdots&,
\end{eqnarray}
where $M_1=\bF$.  We now see that there are three equivalent
complete descriptions of the probability distribution function. The
first one is in terms of binned PDF parameters $B_i$, the second one
is in terms of either moments $M_n$ or central normalized moments
$m_n$, and the third one is in terms of coefficients of Legendre
polynomials $a_\ell$, which are completely given by the moments and
allow one to ``reconstruct'' the actual PDF. Although the above
equations might seem unnecessarily complicated, we note that these are trivially
generated algorithmically to an arbitrary order.

We also note that from the observational perspective, using moments or $a_\ell$
coefficients has a distinct advantage in that it is clear how the noise
and $\bar{F}$ uncertainty enter the measurement.  With respect to the noise,
the measurement of the $i$-th coefficient requires one to measure the
noise-subtracted $i$-th moment. For a purely Gaussian noise, this is
trivial, but any non-Gaussianity will contaminate $i>2$ measurements,
probably progressively so. Therefore, in the presence of noise, it is
unlikely that one can measure, with very high precision more than a
few first coefficients $a_\ell$. On the side of $\bar{F}$, its
determination requires modeling of the unabsorbed continuum level,
which is only possible for spectra of extremely high signal to noise
and sufficient number of unabsorbed pixels \cite{2003MNRAS.342L..79S}.
In our case, any uncertainty in $\bar{F}$
can be propagated in a fully understood manner into the covariance matrix
for the final measurement, since we know exactly how it enters the
measurement of the polynomial coefficients.

Finally, this method makes the complementarity with power spectrum
calculations easier to understand. The power-spectrum is quadratic in the
flux field data and hence the $a_2$ coefficient (or equivalently the second
moments), as a function of the smoothing scale, are completely
determined by the measurement of the power spectrum and the mean
flux. The measurement of $a_3$ as a function of smoothing scale is akin to
a reduced bispectrum, etc.

\section{Demonstration on simulations}

\begin{figure}
\centering
\includegraphics[width=.45\textwidth]{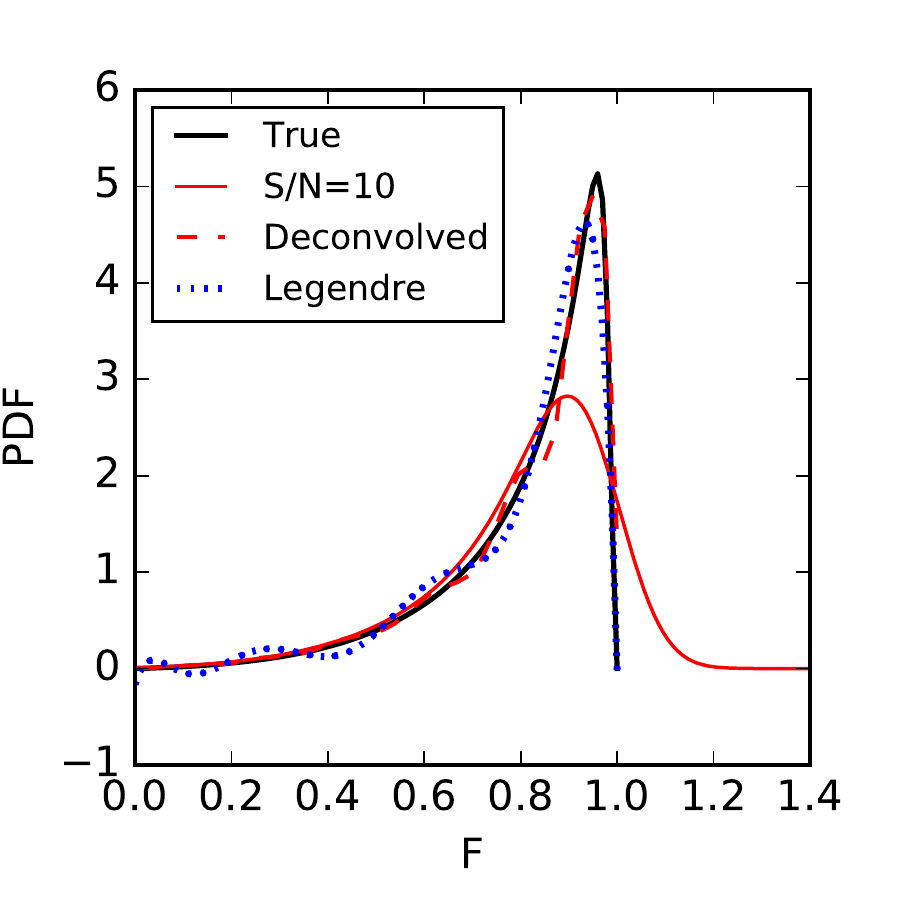}
\includegraphics[width=.45\textwidth]{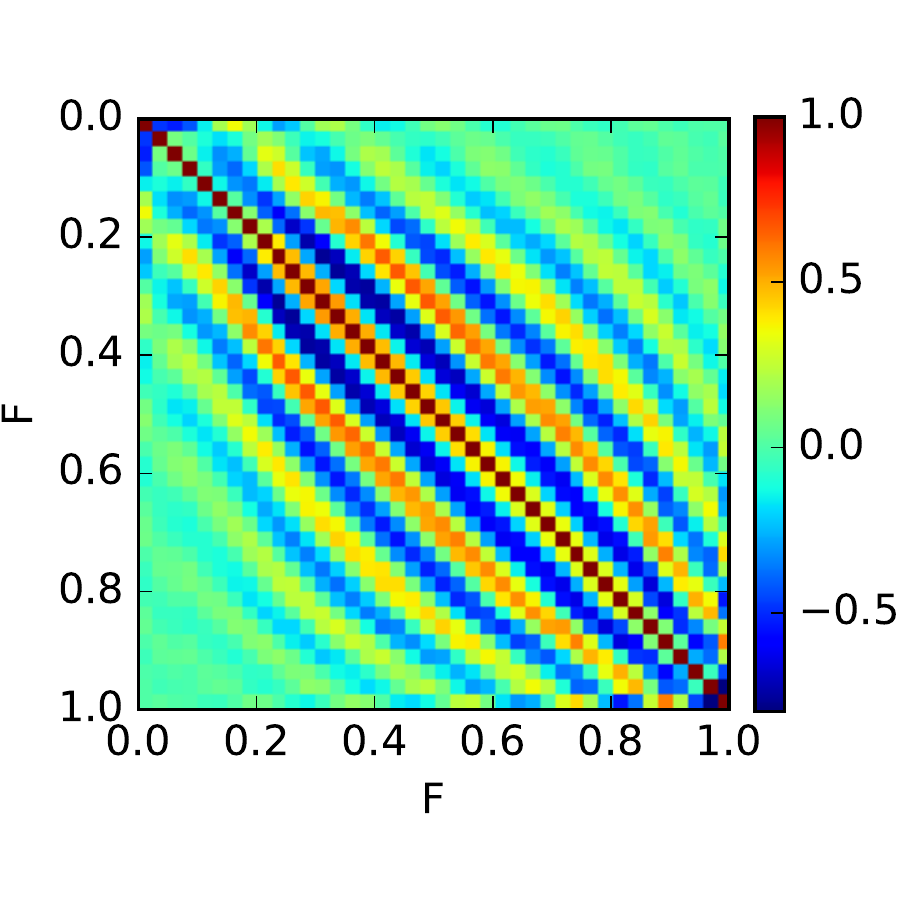}
\caption{True PDF, smoothed with 1 Mpc/h in line-of-sight direction (black solid line), PDF with S/N=10 (red solid line), deconvolved with n=40 (red dashed line) along with the corresponding covariance matrix (right side), and legendre reconstruction with 40 coefficients (dotted blue line).}
\label{fig:PDF_SN10}
\end{figure}

To demonstrate this method numerically, we use the PDF generated from full hydrodynamic
Gadget-3 simulations \cite{Springel2005}, evolved to redshift 2, with a WMAP7
cosmology, where $\Omega_\mathrm{m}=0.275$, $\Omega_\Lambda = 0.725$,
$\Omega_\mathrm{b}=0.046$, $h_0=0.702$, $\sigma_8=0.816$, and
$\mathrm{n}_\mathrm{s}=0.96$. The box size evolved is 80 Mpc/h, with
$N = 2 \times 512^3$ for the number of gas and dark matter particles,
and a Haardt and Madau UV background \cite{1996ApJ...461...20H}. The simple QUICKLYA
option is used for star formation with no feedback.  We generate a flux PDF from
simulated mock Lyman-$\alpha$ forest spectra that have been convolved with the line profile, with the lines of sight on a $512\times 512$ matrix on $1024$ pixel-long skewers. In this demonstration we will consider a rough equivalent of the BOSS
experiment \cite{2013AJ....145...10D} and thus we apply a Gaussian smoothing on the scale of $1h/$Mpc in
the radial direction to simulate the equivalent spectroscopic resolution. This noiseless PDF is pictured in Figure \ref{fig:PDF_SN10} as the black solid line. Since resolution requirements increase with redshift, we use the redshift 2 output for our method demonstration, however we note that our simulation might not fully resolve the extreme fluxes at both ends of the PDF, as noted in convergence studies of \cite{2015MNRAS.446.3697L} and \cite{2017MNRAS.464..897B}. For this example exercise, to overcome the shortcoming of a small box size, we use this PDF as a representative PDF and generate new skewers with pixels drawn independently from this PDF. In order to calculate the covariance matrix for the flux PDF and the subsequent method covariance matrices, we draw points representing 160,000 quasars of 200 pixels each to represent a BOSS-scale experiment  \cite{2013AJ....145...10D}, applying the bootstrap method with 1000 for the number of resamples. 

\begin{figure}
\centering
\includegraphics[width=.45\textwidth]{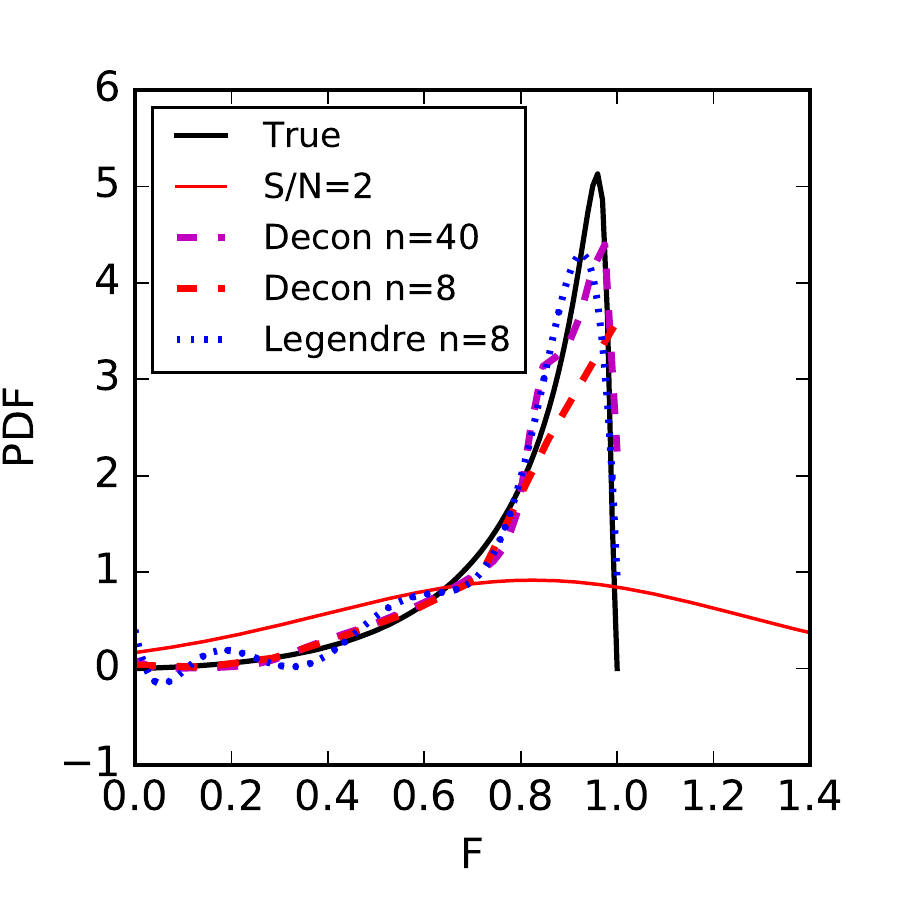}
\includegraphics[width=.45\textwidth]{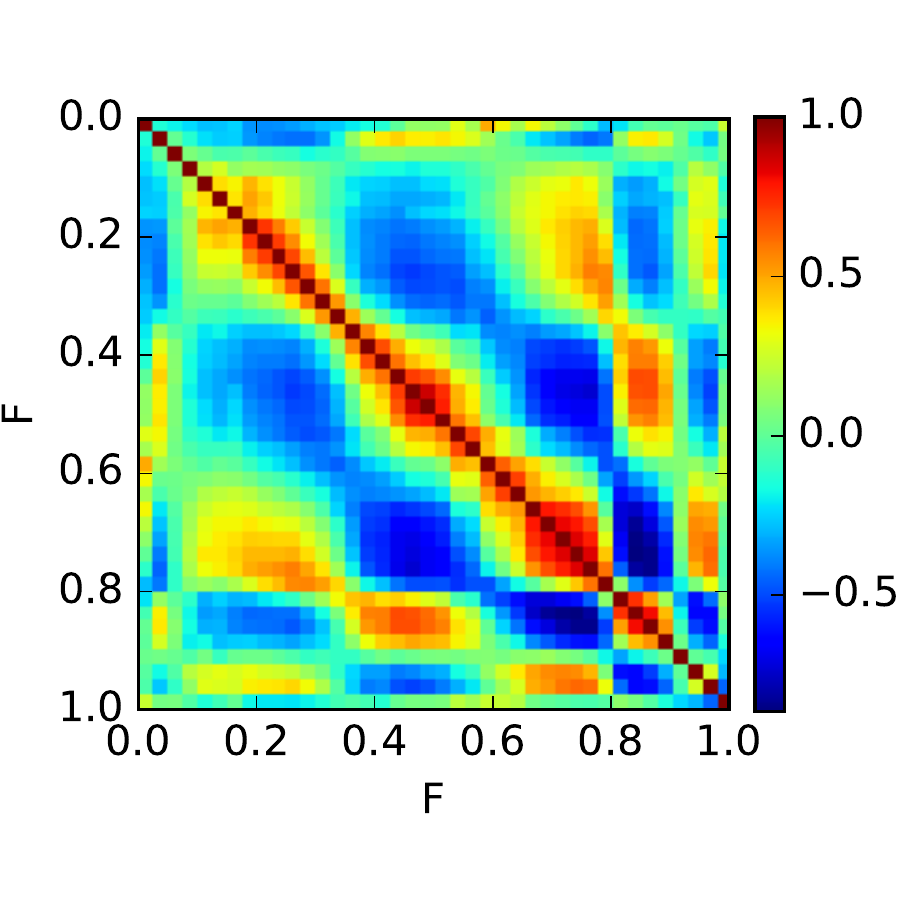}
\caption{True PDF, smoothed with 1 Mpc/h in line-of-sight direction (black solid line), PDF with S/N=2 (red solid line), deconvolved with n=40 (magenta dashed line) along with the corresponding covariance matrix (right side), deconvolved with n=8 (red dashed line), and legendre reconstruction with 8 coefficients (dotted blue line).}
\label{fig:PDF_SN2}
\end{figure}

\subsection{PDF maximum likelihood noise deconvolution}

Next we add varying amount of noise to the data.For the first example, the thin red line in Figure
\ref{fig:PDF_SN10} shows the smoothed binned PDF for a SNR of 10, where we did not
attempt to reconstruct the input PDF. The effect of noise is to
broaden the values of flux beyond the physically allowed range $0-1$ and
demonstrates that a full forward-modeling is required to reconstruct
the input flux PDF.

We first demonstrate a reconstruction of the original PDF using a
  maximum likelihood noise deconvolution method for this toy
  model. We'd like to recover the reconstructed PDF divided into $n$
  bins, where the response of the $n$-th bin to noise spreads the flux
  values into the observed $i$ bins. At fixed noise, the analysis
  simplifies considerably. The observed $i$-th bin of the
  flux PDF can therefore be modeled by the convolution of the
  reconstructed bins $B^r$ with the response $p_i$ of the $n$-th bin
  to noise:
\begin{equation}
B_i^o = B^r * p_i = \sum\limits_n B^r_n p_{i-n}.
\end{equation}

\begin{figure}
\centering
\includegraphics[width=.6\textwidth]{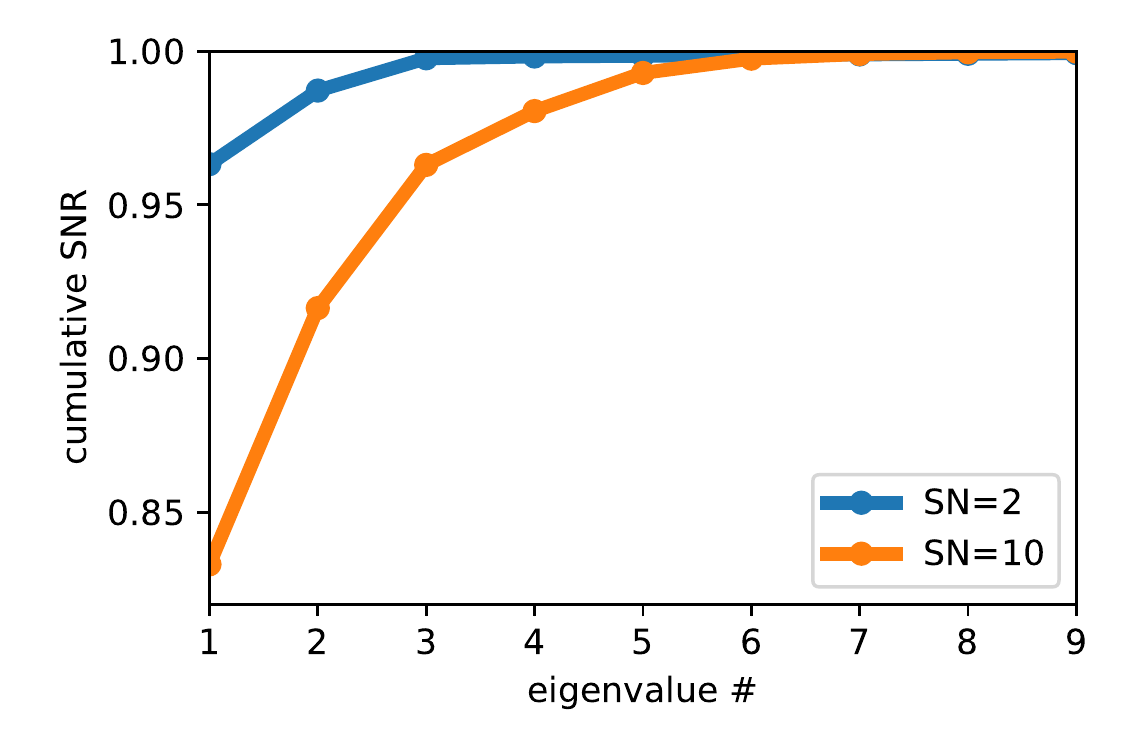}
\caption{The normalized cumulatative signal to noise derived from covariance
  matrices from Figures \ref{fig:PDF_SN10} and \ref{fig:PDF_SN2} as
  described in the text. We infer that all information is compressed in
  $\sim$4 modes for spectral SN$\sim2$ and $\sim8$ modes for spectral SN$\sim5$. }
\label{fig:eig}
\end{figure}

Using the cumulative distribution function for a Gaussian noise distribution:
\begin{equation}
F_n(i) = \frac{1}{2}\left[1+\mathrm{erf}\left(\frac{i-n}{\sqrt{2}\sigma_{\mathrm{noise}}}\right)\right],
\end{equation}
we can model the response of the $n$-th bin to noise as
\begin{equation}
p_{i-n} = F_n(i+1) - F_n(i) = F_i(n)-F_i(n-1),
\end{equation}
where we have used $\mathrm{erf}(-x)=-\mathrm{erf}(x)$. For each bootstrap iteration we therefore can maximize the likelihood by minimizing
\begin{equation}
\chi^2 = \sum\limits_{i} \frac{\left(B_i^o - B^r * p_i \right)^2}{B^r * p_i}
\end{equation}
to arrive at the reconstructed PDF bins $B^r$. The mean of the
reconstructed PDFs is pictured as the red dashed line in Figure
\ref{fig:PDF_SN10} for $n$=40 bins, with the associated covariance
matrix between the bootstrap iterations on the right hand side of the
same figure.   As one can see, the fit to the original PDF is
very good, however the covariance matrix for noise estimation has
complex structure. In addition, as we go to lower SNR the fit worsens,
as is seen in Figure \ref{fig:PDF_SN2}, and the covariance matrix
retains its complex structure. In addition, as we decrease the number
of reconstructed points, the fit becomes drastically worse, even
though the reconstruction is quite good using the same lower number of
Legendre coefficients, as described in the next section. Figure
\ref{fig:PDF_SN2} shows the reconstruction for $n=8$ points, and the
same for 8 measured Legendre coefficients. The $n=40$ bin points shows
a similar fit to the PDF despite five fold increase in the number of
points. In fact the number of well measured modes is very few as we
demonstrate next. We take the 40 bin covariance matrices and
decompose them into eigenvectors and eigenvalues. We then project
the true model into this eigenspace and calculate the signal-to-noise
for each eigenvector, i.e.
\begin{equation}
  {\rm (SNR)}_i^2 = \frac{(\mathbf{v}_i \cdot \mathbf{t})^2}{\lambda_i},
\end{equation}
where  $\lambda_i$ and $v_i$ are the $i$-th eigenvalue and
eigenvector, respectively, and $t$ is the vector containing the true model (i.e. 40
bins representing the PDF).
We then order in decreasing SNR and calculate
the cumulative SNR by summing individual contributions in
quadrature. The results are plotted in Figure \ref{fig:eig}. We
see the expected behavior: out of 40 points, the total SNR in PDF
determination is concentrated in a few linear combinations and the
number of such combinations decreases with spectral SNR. As we will
see later the Legendre polynomial decomposition is only slightly worse
and compresses information into a few well-measured modes.

We can conclude that the deconvolution method  works well for high SNR but does not
  work as well for a given number of points as we decrease the SNR. In
  addition, the covariance matrix of the deconvolved PDF has complex
  structure. 

\subsection{Moments and Legendre coefficients}

\begin{figure}
\centering
\includegraphics[width=.45\textwidth]{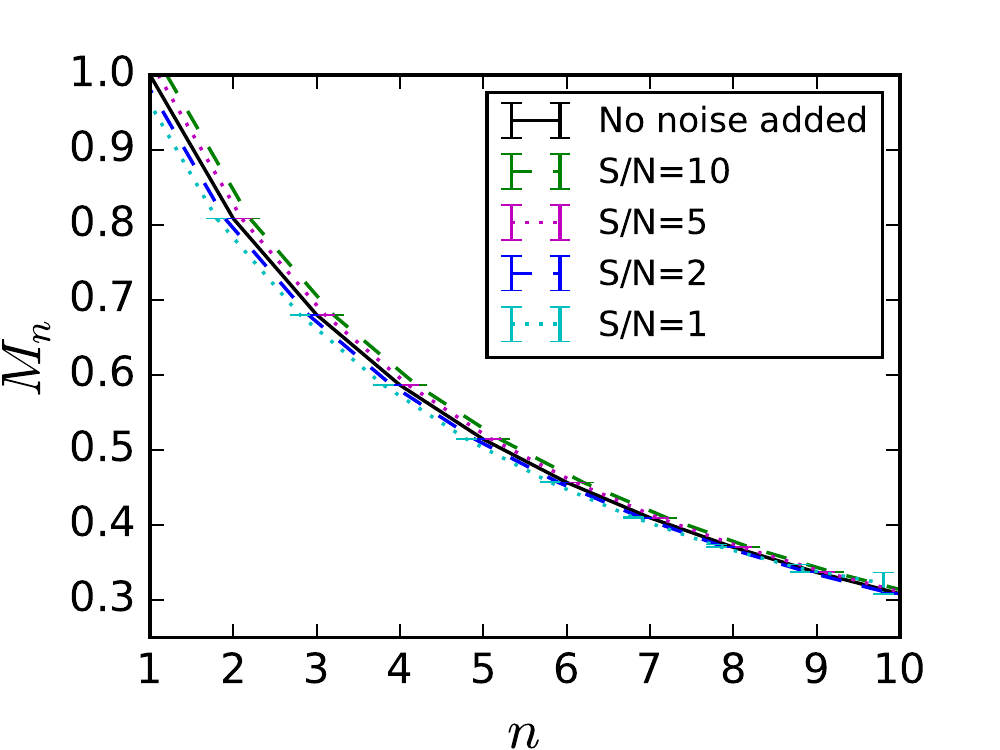}
\includegraphics[width=.45\textwidth]{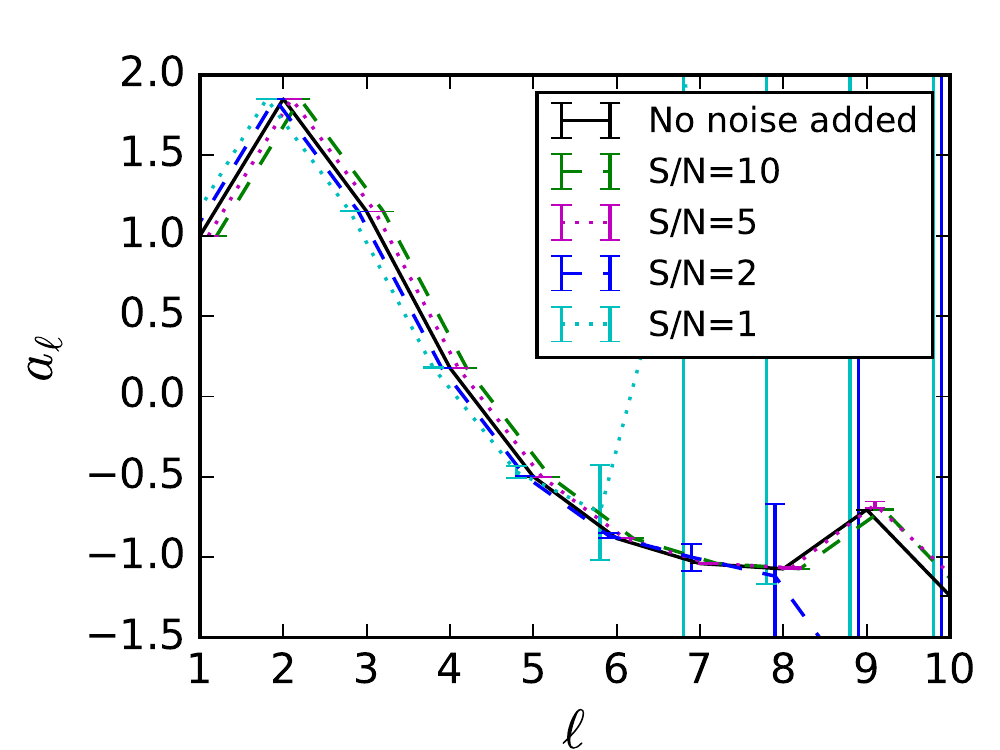}
\caption{Values of $M_n$ in left panel and the reconstructed Legendre polynomial coefficients $a_\ell$  in the right panel after adding Gaussian noise for a signal-to-noise ratio of 1 (cyan dotted line), 2 (blue dashed line), 5 (magenta dotted line), 10 (green dashed line). The no noise added line is represented as a black continuous line. The lines are offset horizontally to better visualize the change in error bars.}
\label{fig:coeffs_err}
\end{figure}

We then calculate moments $M_i$ by measuring raw moments in the data
and then subtracting the noise contribution:
\begin{equation}
   \tilde{M_i} = \frac{1}{N_{\rm points}}\sum_k F_k^i - N_i,
\end{equation}
where $N_2=\sigma^2$, $N_3= 3 \bF \sigma^2$, $N_4=6 \tilde{M_2}\sigma^2+3\sigma^4$, etc. with $\sigma^2$
being the variance of the error. This procedure can be trivially generalized to real
data with varying noise.  This allows us to
reconstruct values of $a_\ell$ and estimate the uncertainty. Since
the coefficients $a_\ell$ are linear in $M_i$, our estimates for $a_\ell$
are unbiased. The results of this exercise are  plotted in Figure \ref{fig:coeffs_err}.
We see that the addition of noise degrades measurements as
expected. However, the effect is considerably more dramatic  for the
$a_\ell$ parameterization than it is for moments. For moments we note a
slow increase in the uncertainties with errors on $M_{10}$ increasing
markedly at low $S/N$ but barely perceptible in others. On the other
hand for $a_\ell$ coefficients, we note a noise-level dependent
``wall'' in $\ell$: for SNR around $1$, there is virtually no
information at $\ell>6$ and for SNR around $2$, the same is true for
$\ell>8$. In other words, the transformation is very successful at
compressing the information content into a few well-measured numbers. 

\begin{figure}
\centering
\includegraphics[width=.45\textwidth]{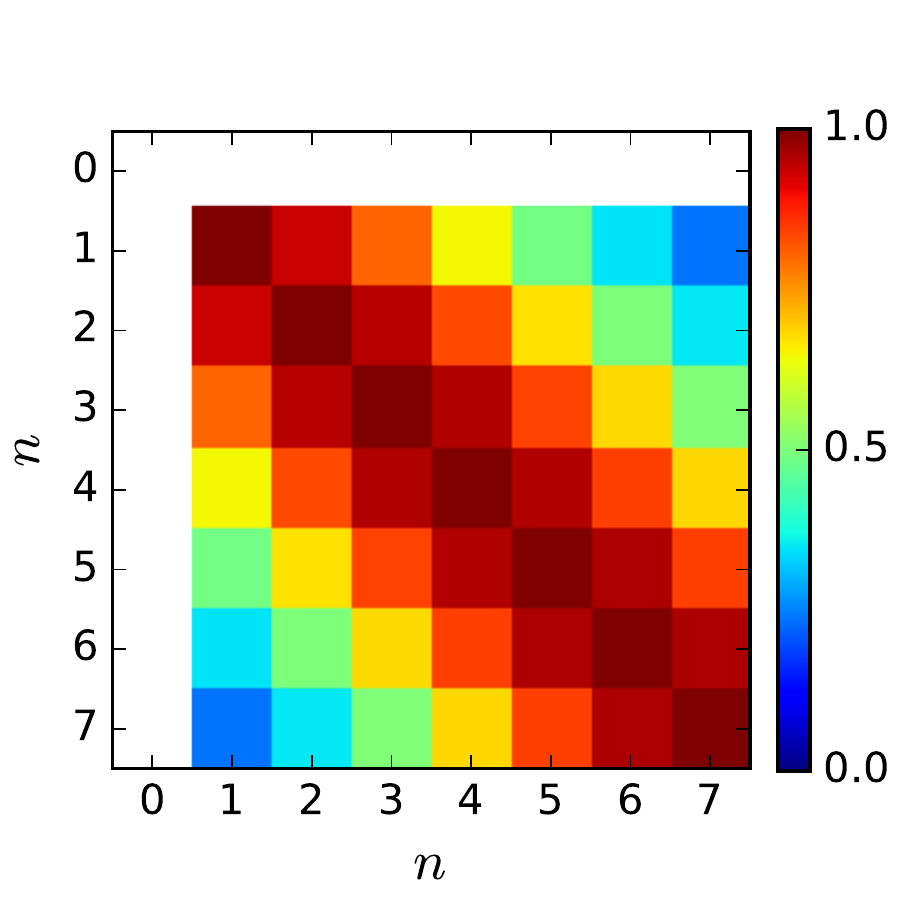}
\includegraphics[width=.45\textwidth]{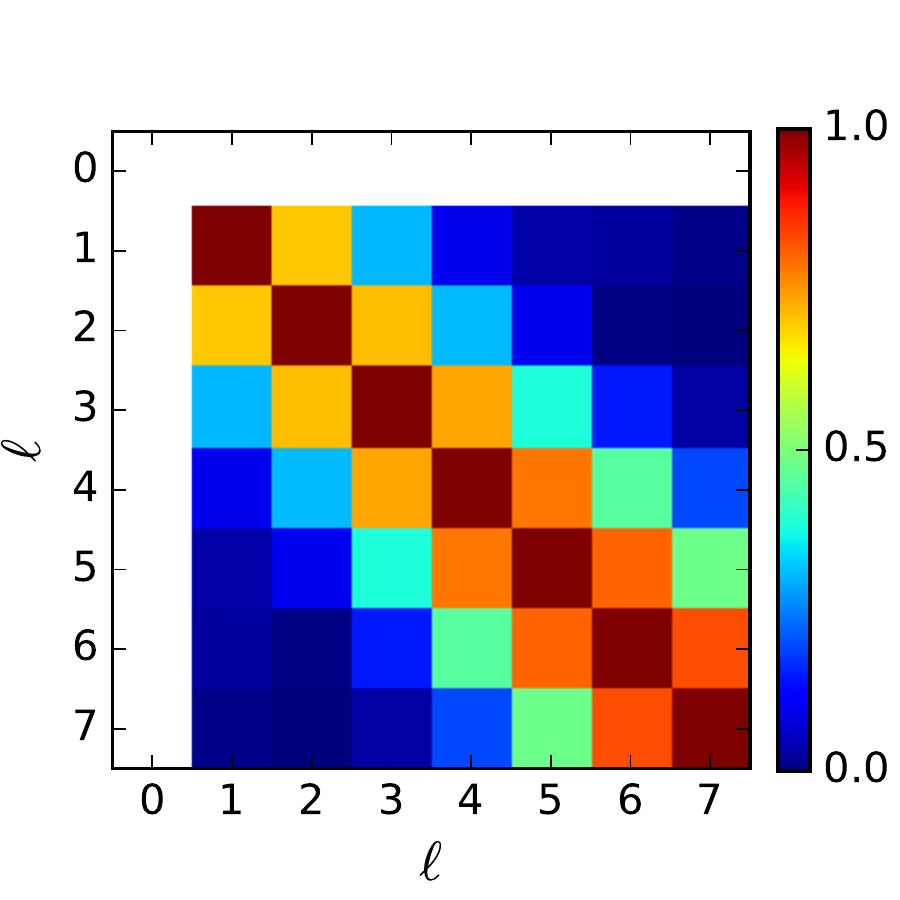}
\caption{Values of the reconstructed $M_n$ covariance matrix in the left panel, and the reconstructed Legendre polynomial coefficients $a_l$ covariance matrix in the right panel, both for a signal-to-noise ratio of 2 (represented as blue dashed line in Figure \ref{fig:coeffs_err}).}
\label{fig:coeffs_cov_err}
\end{figure}

\begin{figure}
\centering
\includegraphics[width=.6\textwidth]{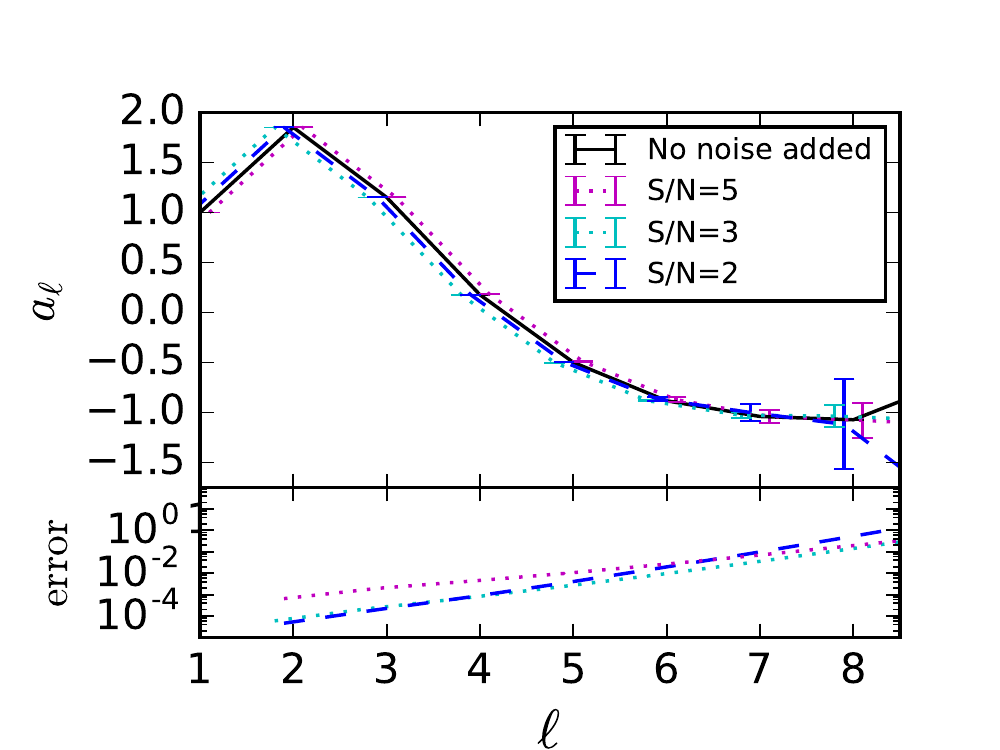}
\caption{Reconstructed $a_l$ coefficients for 160,000
  quasars with a signal-to-noise of 2 (blue dashed line), 5,000
  quasars with a signal-to-noise of 3 (cyan dotted line), and 25
  quasars with a signal-to-noise of 5 (magenta dotted line). The
  bottom panel shows errors. This plot illustrates that higher
  spectral SNR measurements allow determination of a larger number of Legendre coefficients at very modest number of quasars.}
\label{fig:test}
\end{figure}

In Figure \ref{fig:coeffs_cov_err} we show the covariance matrix for
moments and Legendre polynomial coefficients for SNR $\sim2$. We see
that well-determined moments are heavily correlated, but that the
poorly determined ones are virtually uncorrelated. The correlation in
the low $\ell$ modes most likely comes from sample variance which
dominates their determination.

Figure \ref{fig:PDF_SN10} shows the reconstructed PDF against the input
PDF for the most likely solution. This is a sanity check. We see that
SNR $\sim10$ ``reconstructs'' the true PDF very well. 

Since there is a finite number of $a_\ell$ values that can be
  measured well for a given SNR, it is worth exploring the number of
  quasars at higher SNR that would give the same number of
  well-measured Legendre coefficients. In Figure \ref{fig:test} we
  plot the values of $a_\ell$ for SNR=2 for the 160,000 quasars as
  explored earlier and we also plot results for 5000 SNR=3 quasars and
  just 5 SNR=5 quasars. We find that the large number of low SNR
  quasars is only helpful at the first few coefficients and even there
  the required numbers is disproportionally large. 

\section{Conclusions}

The properties of a random field are completely determined by the full
set of $n$-point correlators. Any derived quantity such as PDF can in
principle be derived from them. 

In this paper we have made the link explicit for the Lyman-$\alpha$
forest flux, or any other field, for which the values of the field are
limited to be in the finite range. We have shown that the first $n$
moments of the field uniquely determined the first $n$ $a_\ell$ coefficients in
the Legendre polynomial expansion of the PDF and vice-versa.

Using a toy-example, we argue that instead of measuring PDF, one should measure the
$a_\ell$ coefficients. There are four main reasons for this:
\begin{itemize}
\item In our toy example these coeffients are very
good at sorting the information content available in the PDF into a
few well determined numbers and an infinite number of very poorly
determined numbers. The precise number of well-determined coefficients
is given by the SNR of the available data.

\item Our procedure provides a very clear path to measuring these
  quantities: $a_\ell$ coefficients can be determined from the moments
  of the field for which unbiased estimators in the presence of noise
  are trivially computed.

\item Mean flux $\bar{F}$ is difficult to measure in real data and must be marginalised out assuming a plausible prior. Since
  $\bar{F}$ is just the first moment of the field, this marginalization
  is considerably easier than in the case of fitting a binned flux
  distribution function.

\item If $a_\ell$ coefficients are measured at a number of
smoothing scales, a considerable statistical information is contained in a 2D field
$a_\ell(k_s)$ with $k_s$ being the smoothing wavenumber. In
particular, assuming $\bar{F}$ to be known, the $a_2(k_s)$  contains
the same information as the 1D power spectrum.
\end{itemize}

Before this method can be put in practice, more work needs to be
done. On the observational side, the effects of metal contamination
and damped Lyman-$\alpha$ systems need to be understood. On the
theoretical side, more work needs to be done to establish the
relation between these coefficients and physical properties of the
intergalactic medium, such as its mean flux or temperature-density
relation. We leave this for future work.

Another important result of this paper is how non-linear the
  degradation in ability to characterize the PDF is with spectral SNR. We
  find that, neglecting sample variance, using 25 quasars of SNR=5 yields comparable results to 160,000 quasars of SNR=2. Although the
  latter measures the first few Legendre modes somewhat better, the
  former measures a larger number of the modes. In retrospect, one would expect
  this, because the PDF depends on all moments of the field and the
  measurements of higher modes are of course more sensitive to
  noise. But it is nevertheless quite striking just how large this
  sensitivity is.

\section*{Acknowledgements}

We would like to thank Nishikanta Khandai for providing the Gadget-3 hydrodynamic simulations.

\bibliographystyle{JHEP}
\bibliography{ref,ref2}

\end{document}